# Hybrid density functional theory calculations on phonons in LaCoO$_3$

Denis Gryaznov,*[1,3] Robert A. Evarestov,[2] and Joachim Maier[1]

[1]*Max Planck Institute for Solid State Research, Heisenbergstr. 1, D-70569 Stuttgart, Germany*
[2] *Department of Quantum Chemistry, St. Petersburg University, Universitetsky Prosp. 26, 198504 St. Petergof, Russia*
[3] *Institute for Solid State Physics, Kengaraga 8, LV-1063 Riga, Latvia*

Phonon frequencies at Γ-point in non-magnetic rhombohedral phase of LaCoO$_3$ were calculated using density functional theory (DFT) with hybrid exchange correlation functional PBE0. The calculations involved a comparison of results for two types of basis functions commonly used in *ab intiio* calculations, namely the plane wave (PW) approach and linear combination of atomic orbitals (LCAO), as implemented in VASP and CRYSTAL computer codes, respectively. A good qualitative, but also within an error margin of less than 30%, a quantitative agreement was observed not only between the two formalisms but also between theoretical and experimental phonon frequency predictions. It is concluded that the hybrid PBE0 functional is able to predict correctly the phonon properties in LaCoO$_3$.

From the application point of view, it should be mentioned that cobaltite based electrode materials are extensively studied for oxygen permeation membranes[1] and fuel cells[2]. LaCoO$_3$ is a strongly correlated material with a Co-O covalent bonding. LaCoO$_3$ attracts great interest from researchers due to its complex magnetic behaviour and interesting phase diagram. One of the key issues to be solved concerns the magnetic phase transition from the low spin state (LS: $t_{2g}^6 e_g^0 \rightarrow a_{1g}^2 e_g^4 e_g^0$, i.e. S = 0, here splitting of 3d-levels of Co in cubic and rhombohedral crystal fields, respectively, is shown) to excited high spin state (HS: $t_{2g}^4 e_g^2 \rightarrow a_{1g}^2 e_g^2 e_g^2$, i.e. S = 2) or intermediate spin state (IS: $t_{2g}^5 e_g^1 \rightarrow a_{1g}^2 e_g^3 e_g^1$, S = 1)[3-4]. Also, the semiconductor-metal transition at ~500 K is still not properly understood[5]. As the standard DFT exchange-correlation functionals predict LaCoO$_3$ magnetic and metallic at low temperatures opposite to experiments (see discussion below), obviously a careful theoretical treatment is required. The majority of publications discussing the *ab initio* calculation results for LaCoO$_3$ are based on the so-called LDA(GGA)+U method, which was shown to reproduce correctly a non-magnetic behavior of LaCoO$_3$ at low temperatures[6-7]. A commonly used approach is to find the U-parameter by fitting materials properties to experimental values[8]. The U-values so far discussed in the literature for LaCoO$_3$ vary in the range between ~3 and ~8 eV depending on the basis and Hamiltonian. These techniques demonstrated a LS ground state and an IS state to be the lowest excited state and, thus, proposed an explanation to controversy on the magnetic phase transition appearing at temperatures close to 90 K[6].

Recent quantum chemical non-periodic (embedded cluster) LCAO calculations did not confirm the stabilization of IS state against the HS state[9]. In these calculations the restricted Hartree-Fock method and atom-centred Gaussian-type orbitals were applied. However, the LS state was found to be stabilized only if dynamical electron correlation effects were included.

It is worth mentioning, that the magnetic phase transitions discussed are thermally induced. The temperature effects could be included by calculating the phonon frequencies as the pre-requisite to free energies.

In the present study the two different periodic approaches (namely, LCAO and PW) are applied to calculate the phonon frequencies for the LS state of LaCoO$_3$ within the same hybrid exchange-correlation functional, namely PBE0[10]. The results are also compared with recently published LDA+U study of Laref and Luo[11]. Unlike the LDA(GGA)+U method, the hybrid PBE0 functional does not require any adjustable parameters and has been deduced on the basis of pure theoretical grounds[10]. This functional has shown to properly reproduce the properties of many strongly correlated systems (see, for example, Ref. 12).

The calculations of LaCoO$_3$ phonon frequencies were based on the following procedure. First, the equilibrium geometry was found by optimizing the structure parameters (see discussion below) for the $D_{3d}^6 (R\bar{3}c)$ space group. The rhombohedral cell parameters were compared with experimental data. Second, the phonon frequencies were obtained within the frozen phonon method[13-14] at the fixed equilibrium structure parameters.

The phonon frequencies in the LCAO basis were calculated in Crystal09[15]. The Gaussian basis set and pseudopotentials (PP) of free atoms for La and Co were taken from the PPs library of the Stuttgart/Cologne group[16]. The relativistic PP for La included 46 electrons and for Co 10 electrons. In order to avoid spurious interactions between the diffuse functions and the core functions of neighboring atoms, the basis set diffuse exponents smaller than 0.1 (in Bohr$^{-2}$) for La and Co were removed. The all electron basis set 8-411G* from Ref. 17 was used for O atom for this non-optimized basis set (hereafter, BS1). In the optimized basis set (hereafter, BS2) the exponents of non-contracted basis functions were optimized by the method of conjugate directions[18] as implemented in the OPTBAS code[19]. The role of BS optimization is discussed by comparing phonon frequencies and bulk properties for both the BSs. The Monkhorst-Pack scheme for 8x8x8 k-point mesh in the Brillouin zone (BZ) was applied together with the tolerances 7, 7, 7, 7, 14 for the Coulomb and exchange integrals calculations. The tolerances were increased to 10, 10, 10, 10, 16 for the phonon frequency calculations. Furthermore, the forces for the self consistent cycles were optimized until the energy difference reached $10^{-6}$ eV for the lattice structure optimization and $10^{-8}$ eV for the phonon frequency calculations.

The phonon frequencies in the plane wave (PW) basis were calculated in VASP code[20-21], which represents an effective tool for solid state calculations based on the ultra-soft pseudopotentials approach and PWs. We used the projector augmented wave (PAW) method[22] and scalar-relativistic PPs substituting for 46 core electrons on La atom, 18 core electrons on Co atoms and 2 core electrons on O atom. The plane wave cut-off energy was fixed at 520 eV for the geometry optimization and increased to 600 eV for the phonon frequency calculations. The tolerance for forces calculations and the Monkhorst-Pack scheme were the same as in the Crystal calculations. The electron occupancies were determined with the Gaussian method using a smearing parameter of 0.1 eV.

Table 1. A comparison of rhombohedral lattice constant $a$ (in Å), angle α (in degree), the O position free parameter $u$ (in fraction of the lattice constant), effective charge Q (in $e^-$) of La, Co and O atoms, and band gap $E_g$ (in eV) for different methods used in the present study and the experiment.

| Method | $a$ | α | $u$ | $Q_{Co}$ | $Q_{La}$ | $Q_O$ | $E_g$ |
|---|---|---|---|---|---|---|---|
| LCAO PBE (BS2) | 5.41 | 61.48 | 0.193 | 1.30 | 2.66 | -1.32 | Metal |
| LCAO PBE0 (BS1) | 5.34 | 60.86 | 0.209 | 0.95 | 2.76 | -1.24 | 3.19 |
| LCAO PBE0 (BS2) | 5.36 | 60.83 | 0.206 | 1.40 | 2.73 | -1.38 | 3.14 |
| PW PBE | 5.37 | 61.53 | 0.186 | 1.40 | 2.09 | -1.16 | Metal |
| PW PBE0 | 5.33 | 61.01 | 0.195 | 1.62 | 2.23 | -1.28 | 2.50 |
| LDA+U [11] | 5.34 | 60.91 | 0.195 | - | - | - | ~1.8 |
| Exp | 5.34[a] | 60.99[a] | 0.198[a] | 3[b] | 3[b] | -2[b] | 0.6[c] |

[a] at 4 K Ref. [23]; [b] formal charge; [c] Ref. [29]

LaCoO$_3$ has a rhombohedrally distorted perovskite structure known to exist within a broad temperature range[23-25].

Table 1 presents the results of *ab initio* PBE and PBE0 calculations for structure parameters, effective charges (due to Mulliken analysis in Crystal[26]) and Bader analysis[27-28] in VASP) and the electronic band gap. The results are also compared to experimental data. It is worth mentioning that the standard DFT exchange-correlation functional (PBE) predicts LaCoO$_3$ to be metallic using the LCAO as well as PW approach. Contrary, the hybrid PBE0 functional suggests that LaCoO$_3$ is insulating with a band gap $E_g$ of 2.5 eV and 3.14 eV in the PW and LCAO (BS2) calculations, respectively. Such values of $E_g$ are higher than those known from the LDA(GGA)+U calculations: 1.0 eV[7], 1.43 eV[8], and 2.0 eV[6]. It can, at least partly, be explained by the exact exchange part of the PBE0 functional which tends to overestimate $E_g$. The ultraviolet photoemission spectroscopy if combined with the bremsstralung isochromat spectroscopy[29] can provide us with the measured value of $E_g$. However, the value of $E_g$ suggested in Ref. 29 is smaller than our value found using the hybrid functionals (table 1). The interpretation of measured spectra is not simple, and Saitoh et. al[30] also deduced the value of 2 eV from Ref. 29. It is obvious that LaCoO$_3$ is a non-magnetic insulator at low temperatures which is correctly reproduced within our ab initio calculations. The lattice parameters are very well reproducible using the PBE0 functional (the errors of the order 1% or less). The standard PBE functional slightly overestimates the rhombohedral lattice constant $a$ and angle α. The coordinates of O atoms in the $D_{3d}^6$ space group are characterized by the free parameter $u$ (see also table 2). The value of $u$ is underestimated by the PBE functional. This is why; the PBE functional was excluded from our following analysis of calculated phonon frequencies. There is, however, a very good agreement on the parameter $u$ between the LDA+U study[11] and present PBE0 calculations in the PW basis. In contrast, the LCAO calculations suggest slightly higher values of $u$.

Table 2 gives the symmetry of phonons in two LaCoO$_3$ structures: the cubic $O_h^1$ and rhombohedral $D_{3d}^6$. The symmetry is defined by the corresponding space group irreducbile representations (irreps) induced from atom site symmetry group irreps corresponding to the atomic displacement x,y,z. In particular, the O atom in the d-position in the cubic structure has D$_{4h}$ symmetry. However, its local symmetry in $D_{3d}^6$ is reduced to C$_2$ and coordinates are determined by free parameter $u$ which defines the rhombohedral angle. One can see that the O and Co atoms contribute to the a$_{1u}$ irrep, the O and La atoms to the a$_{2g}$ and e$_g$ irreps, all three atoms - to the a$_{2u}$ and e$_u$ irreps, and there is one a$_{1g}$ irrep determined by the O atom only. Table 2 also shows that the t$_{1u}$ irrep of La in the cubic crystal is splitted into a$_{2u}$ and e$_u$ in the rhombohedral crystal whereas the a$_{2g}$ and e$_g$ are appearing by the projection of R-point of the cubic crystal onto Γ-point in the rhombohedral crystal. On the other hand, the t$_{1u}$ irrep of Co in the cubic crystal is splitted into a$_{1u}$ and e$_u$ in the rhombohedral crystal, and $R_5^+$ in the cubic crystal is splitted into $a_{1u}$ and $e_u$ in the rhombohedral crystal.

A crystal with rhombohedral structure such as LaCoO$_3$ has the following set of optical phonon modes at Γ-point: $a_{1g}+4e_g+6e_u+4a_{2u}+3a_{2g}+2a_{1u}$ (table 2). All such calculated modes except for $3a_{2g}$ and $2a_{1u}$ are given in table 3 for the PBE0 functional in two codes together with the experimental frequencies $\nu_{exp}$ from the literature. The latter two modes are silent and are not visible in optical experiments. Owing to errors in the numerical calculation of total energy second derivative over the displacements, two acoustic modes are not exactly zero but small imaginary numbers. In table 3 the relative errors with respect to $\nu_{exp}$ are defined as $|\nu_{exp} - \nu_{th}|/\nu_{exp} \cdot 100\%$, where $\nu_{th}$ is the calculated frequency. The Raman active phonon modes calculated in both Crystal and VASP are close to $\nu_{exp}$ from Ref. 31. The LCAO basis optimization (BS2 vs BS1) improves the Raman active $\nu_{th}$, suggesting the largest error of 21% for the lowest $e_g$ frequency with respect to the experiment. The PW PBE0 calculations show an error of 34% for the same mode. Additionally, the results from the LDA+U study[11] are given in table 3. The lowest $e_g$ frequency is better reproduced in their study whereas the highest $e_g$ frequency is obtained with the error 23% opposite to 5% in our PBE0 calculations. In order to properly discuss the IR modes; one has to keep in mind, that the eight IR active optical modes in the rhombohedral crystal are splitted into $3(a_{2u}+e_u)$ combinations and $2e_u$ modes. The three pairs are denoted in the literature as external, bending and stretching modes from low to high frequencies. The experimental phonon modes from the reflectivity spectra of Yamaguchi et. al[32] and Tajima et. al[33] showed three major peaks. The low temperature measurements of Yamaguchi et. al could also identify weak peaks from the two $e_u$ modes whereas Tajima et. al suggested partial assignment of the $3(a_{2u}+e_u)$ peaks: the peaks were separated into pairs, however, a complete assignment of $a_{2u}$ and $e_u$ is not clear. On the basis of the fact that the $a_{2u}$ mode in all the calculations has lower frequency, we divided 6 frequencies from table 1 in Ref. 33 into $a_{2u}$ and $e_u$ modes and compared them to the calculated frequencies. For the $3(a_{2u}+e_u)$ pairs the $a_{2u}$ mode is well comparable to the experiment with the maximum error 13% for the BS1 basis in LCAO PBE0 calculations. The basis optimization reduced this error to 2% only (see BS2 in table 3). Contrary, the $e_u$ modes are worse comparable to the experiment (the maximum error 20% for the lowest $e_u$ frequency). Similar results were also observed for the PW PBE0 calculations: the error 26% for the lowest $e_u$ mode. Such high errors for the $e_u$ modes may be also related to limitations of the experiment in Ref. 33 due to only partial symmetry assignment. The other two weak $e_u$ modes are perfectly reproduced in the PW PBE0 calculation (errors of 2% and 7%, respectively) and a bit worse in the LCAO PBE0 calculation for BS2 (errors 13% and 5%). However, the IR frequencies are poorly reproduced in the LDA+U calculation[11] leading to errors as high as 60% for majority modes.

Table 2. Atomic Wyckoff position, site symmetry and simple induced representations (irreps) of the $O_h^1$ and $D_{3d}^6$ space groups.

| $O_h^1 (Pm\bar{3}m)$ | | | | $D_{3d}^6 (R\bar{3}c)$ | | |
|---|---|---|---|---|---|---|
| Wyckoff position | irreps | | | Wyckoff position | irreps | |
| | Γ | R | | | Γ | |
| La - a(0,0,0) $O_h$, $t_{1u}(x,y,z)$ | 4$^-$($t_{1u}$) | 4$^-$ | | La - 2a $\left(\frac{1}{4},\frac{1}{4},\frac{1}{4}\right)$ $C_{3v}$, $a_2(z)$ $e(x,y)$ | 2$^+$($a_{2g}$) 2$^-$($a_{2u}$) 3$^+$($e_g$) 3$^-$($e_u$) | |
| Co - b $\left(\frac{1}{2},\frac{1}{2},\frac{1}{2}\right)$ $O_h$, $t_{1u}(x,y,z)$ | 4$^-$($t_{1u}$) | 5$^+$ | | Co - 2b (0,0,0) $C_{3i}$, $a_u(z)$ $e_u(x,y)$ | 1$^-$($a_{1u}$) 2$^-$($a_{2u}$) 2x3$^-$($e_u$) | |
| O - d $\left(0,0,\frac{1}{2}\right)$ $D_{4h}$, $a_{2u}(z)$ $e_u(x,y)$ | 4$^-$($t_{1u}$) 4$^-$($t_{1u}$) 5$^-$($t_{2u}$) | 1$^+$3$^+$ 4$^+$5$^+$ | | O - 6e $\left(-u, u+\frac{1}{2}, \frac{3}{4}\right)$ $C_2$, $a(z)$ $2b(x,y)$ | 1$^+$($a_{1g}$) 1$^-$($a_{1u}$) 3$^+$($e_g$) 3$^-$($e_u$) 2x2$^+$($a_{2g}$) 2x2$^-$($a_{2u}$) 2x3$^+$($e_g$) 2x3$^-$($e_u$) | |

Table 3. Phonon symmetry (only active modes) and frequencies $\nu_{th}$ in cm$^{-1}$ calculated in VASP (PW) and Crystal (LCAO) with the PBE0 exchange-correlation functional. The experimental Raman and Infrared (IR) frequencies $\nu_{exp}$ from Ref. 31 and Refs. 32-33, respectively, and the phonon modes from the LDA+U study are also given. The values in the parenthesis represent the relative errors with respect to the experimental values in %.

| | Raman active | | | | | Acoustic | | IR active | | | | | | | |
|---|---|---|---|---|---|---|---|---|---|---|---|---|---|---|---|
| Phonon Symmetry | $e_g$ | $e_g$ | $a_{1g}$ | $e_g$ | $e_g$ | $e_u$ | $a_{2u}$ | $a_{2u}$ | $e_u$ | $e_u$ | $a_{2u}$ | $e_u$ | $e_u$ | $a_{2u}$ | $e_u$ |
| $\nu_{exp}$ | 86 | 172 | 261 | 432 | 584 | 0 | 0 | 177$^a$ 180$^b$ 174+240$^b$ | 242$^a$ | 315$^a$ 330$^b$ 328+411$^b$ | | 411$^a$ | 540$^a$ 600$^b$ 550+582$^b$ | | |
| LCAO PBE0 (BS1) | 108 (26) | 193 (12) | 238 (9) | 452 (5) | 646 (11) | 5.1i | 0.1i | 196 (13) | 209 (13) | 280 (16) | 327 (0) | 364 (11) | 433 (5) | 554 (1) | 565 (3) |
| LCAO PBE0 (BS2) | 104 (21) | 178 (3) | 257 (1) | 448 (3) | 631 (8) | 0.7i | 0.0 | 171 (2) | 193 (20) | 273 (13) | 329 (0) | 362 (12) | 433 (5) | 536 (2) | 551 (5) |
| PW PBE0 | 57 (34) | 178 (3) | 280 (7) | 438 (1) | 613 (5) | 4i | 5i | 151 (13) | 177 (26) | 247 (2) | 305 (7) | 351 (15) | 441 (7) | 513 (7) | 532 (9) |
| PW LDA+U$^c$ | 78 (9) | 174 (1) | 253 (3) | 409 (5) | 448 (23) | - | - | 72 (59) | 53 (78) | 79 (67) | 532 (62) | 111 (73) | 158 (62) | 652 (18) | 412 (29) |

$^a$ Ref. [32]; $^b$ Ref. [33]; $^c$ Ref. [11]

In summary, several DFT approaches available to calculate the phonon frequencies were compared for $LaCoO_3$. We have clearly shown that the PBE0 exchange-correlation functional is able to predict the phonon frequencies in $LaCoO_3$, independently of the basis choice (LCAO and PW). Also, we have observed a trend that the highest errors in the phonon frequencies calculated by the frozen phonon method appear for the lowest frequencies; the errors decrease for the intermediate frequencies, and slightly increase for the highest frequencies. We believe that this is a general trend, which must be taken in the analysis of calculated phonon frequencies. The results as obtained in the present study are very important for future analysis of spin state transitions using phonon modes and calculating thermodynamic properties in $LaCoO_3$.

This study was partly supported by EC NASA – OTM project. Authors are grateful to Prof. Manuel Cardona and Dr. Natalya Kovaleva for helpful discussions, Prof. Eugene Kotomin for carefully reading the manuscript and Maxim Losev for the support with the OPTBAS package.

* d.gryaznov@fkf.mpg.de